\documentclass[twocolumn,amsmath,amssymb,a4paper,prb,superscriptaddress,floatfix]{revtex4-1}
\usepackage[dvipdfmx]{graphicx}
\usepackage{natbib}
\usepackage{multirow}
\usepackage{amsmath}
\usepackage{bm}
\usepackage{mathrsfs}
\usepackage{hyperref}
\usepackage{physics}
\usepackage{url}
\usepackage{color}
\usepackage{ulem}
\begin{document}

\title{LO-mode phonon of KCl and NaCl at 300 K by inelastic X ray scattering
  measurements and first principles calculations}

\author{Atsushi Togo}
\email{togo.atsushi@gmail.com}
\affiliation{Research and Services Division of
  Materials Data and Integrated System, National Institute for Materials
  Science, Tsukuba, Ibaraki 305-0047, Japan}
\affiliation{Center for Elements Strategy Initiative for Structural
  Materials, Kyoto University, Sakyo, Kyoto 606-8501, Japan}

\author{Hiroyuki Hayashi}
\affiliation{Department of Materials Science and
  Engineering, Kyoto University, Sakyo, Kyoto 606-8501, Japan}

\author{Terumasa Tadano}
\affiliation{Research Center for Magnetic and Spintronic Materials,
  National Institute for Materials Science, Tsukuba, Ibaraki 305-0047,
  Japan}

\author{Satoshi Tsutsui}
\affiliation{Japan Synchrotron Radiation Research Institute,
  Sayo-cho, Hyogo 679-5198, Japan}
\affiliation{{Institute of Quantum Beam Science, Graduate School of Science and
      Engineering, Ibaraki University, Hitachi, Ibaraki 316-8511, Japan}}

\author{Isao Tanaka}
\affiliation{Center for Elements Strategy Initiative for Structural
  Materials, Kyoto University, Sakyo, Kyoto 606-8501, Japan}
\affiliation{Department of Materials Science and
  Engineering, Kyoto University, Sakyo, Kyoto 606-8501, Japan}
\affiliation{Nanostructures Research Laboratory, Japan Fine Ceramics
  Center, Atsuta, Nagoya 456-8587, Japan}

\begin{abstract}
  Longitudinal-optical (LO) mode phonon branches of KCl and NaCl were measured
  using inelastic X-ray scattering (IXS) at 300 K and calculated by the
  first-principles phonon calculation with the stochastic self-consistent
  harmonic approximation. Spectral shapes of the IXS measurements and calculated
  spectral functions agreed well. We analyzed the calculated spectral functions
  that provide higher resolutions of the spectra than the IXS measurements. Due
  to strong anharmonicity, the spectral functions of these phonon branches have
  several peaks and the LO modes along $\Gamma$--L paths are disconnected.
\end{abstract}
\maketitle

\section{Introduction}
\label{sec:introduction}
Phonon is a picture to represent collective vibrations of atoms in crystal, and
known to play important roles in determining a variety of
crystal properties such as heat capacity, thermal expansion, and thermal
conductivity. Phonons in Brillouin zones of crystals have been measured using
inelastic neutron and X-ray scatterings, and we can find phonon band structures
of those crystals in literature. In the last decades, computational advances to
solve electronic Sch\"{o}dinger equation within the density functional
theory~\cite{Hohenberg-DFT-1964, Kohn-DFT-1965} (DFT) enabled us to predict phonon
properties.\cite{Kunc-phonon-1982, Giannozzi-DFPT-1991, Gonze-1994, Gonze-1997,
  Parlinski-phonon-1997}. Due to its strong predictability, nowadays, phonon
calculation is applied to studies in various scientific fields as an essential
tool.

A picture of phonons is typically introduced by a simple coordinate
transformation of basis to represent crystal potential from space of atomic
displacements to that of collective atomic
displacements.\cite{Ziman-electrons-phonons, Thermodynamics-of-crystals} In many
crystals at modest temperatures, phonon properties of crystals are often well
reproduced in perturbation theory, where Taylor expansion of crystal potential
with respect to atomic displacements is truncated at lowest order terms as a
good approximation. As a result, the computational procedure of the phonon
calculation becomes simple and the practical application is made systematic.
This is one of the reasons that the phonon calculation has become popular in
scientific research. Majority of reported phonon calculations are limited to the
harmonic approximation since it often satisfies our requirements for our
researches and in practice, it is much less computationally demanding than a
beyond-harmonic treatment.

Crystals that are difficult to apply straightforward perturbation theory may be
categorized as anharmonic crystals. Anharmonicity is a ubiquitous phenomenon,
that are related to properties of materials such as thermal expansion, lattice
thermal conductivity, and structural phase
transition.\cite{Thermodynamics-of-crystals,
  Solid-state-physics-Ashcroft-Mermin} When we study anharmonic crystals, it is
important to know the phonon spectral shapes, for which the harmonic
approximation is insufficient. Recent progress of computational methodologies in
anharmonic phonon calculations~\cite{Errea-SSCHA-2013, Hellman-TDEP-2013,
  Tadano-2015} has enabled us to simulate phonon spectra of anharmonic crystals.
However comparison of detailed phonon spectral shapes between experiments and
calculations is non-trivial since high-resolution experimental measurements are
limited.

In this study, for the purpose of the comparison, we measured vibrational
spectra of rocksalt-type KCl and NaCl using inelastic X-ray scattering (IXS)
installed at BL35XU of SPring-8. These crystals were chosen among simple
crystals calculated for our previous study~\cite{Seko-LTC-2015} since their
longitudinal optical (LO) modes were expected to exhibit larger linewidths than
the resolution of the IXS instrument, and these crystals used for the
measurements were easily obtained. {An advantage of using IXS
    over inelastic neutron scattering (INS) is to allow the direct comparison
    between experimental and calculated spectral shapes at each q-point
    since energy and momentum transfers as coordinates in IXS measurement are
    uncorrelated in IXS.}

\begin{figure*}[ht]
  \begin{center}
    \includegraphics[width=0.95\linewidth]{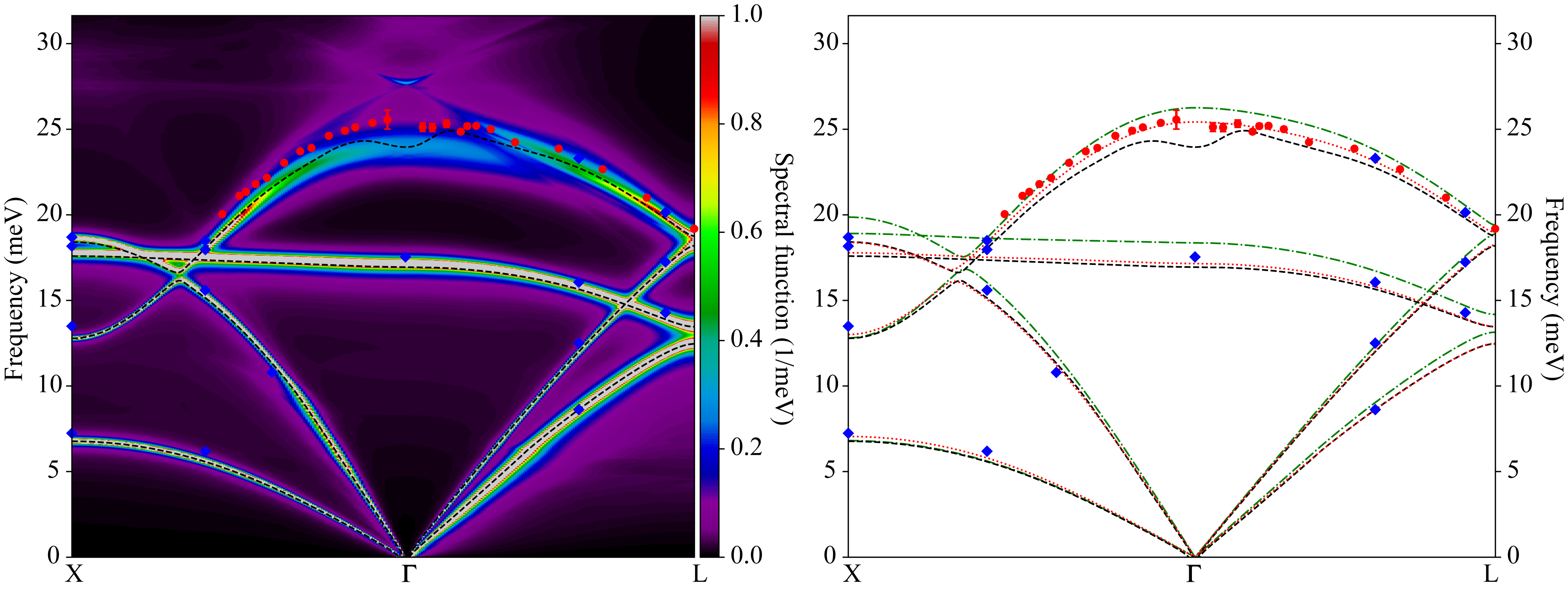} \caption{ (Left panel)
      Phonon spectral function of KCl, $\sum_\nu A_{\mathbf{q}\nu}(\omega)$,
      along X--$\Gamma$--L path at 300K. (Right panel) Phonon band structure of
      KCl along X--$\Gamma$--L path at 300K. The (green) dashed-dotted curve
      shows the renormalized frequencies ($\Omega_{\mathbf{q}\nu}$), the (black)
      dashed curve depicts the renormalized frequencies shifted by the real part
      of the self-energy ($\Delta_{\mathbf{q}\nu}(\Omega_{\mathbf{q}\nu})$) and
      the (red) dotted curve shows the harmonic frequencies. The filled circle
      and diamond symbols show peak positions of the IXS spectra by our
      measurement and the INS measurement by Raunio and
      Almqvist~\cite{Raunio-KCl-1969}, respectively. \label{fig:KCl-sf} }
  \end{center}
\end{figure*}

\begin{figure*}[ht]
  \begin{center}
    \includegraphics[width=0.95\linewidth]{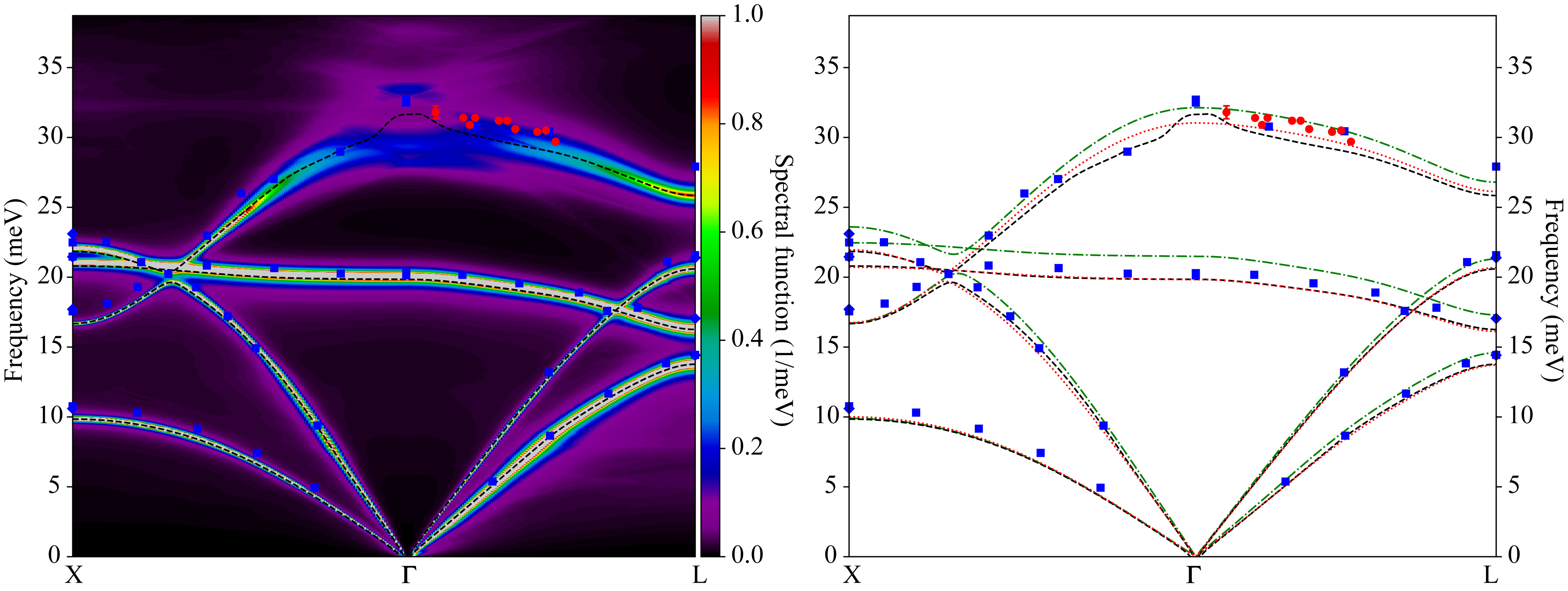}
    \caption{(Left panel) Phonon spectral function of NaCl, $\sum_\nu
        A_{\mathbf{q}\nu}(\omega)$, along X--$\Gamma$--L path at 300K. (Right
      panel) Phonon band structure of KCl along X--$\Gamma$--L path at 300K.
      The (green) dashed-dotted curve shows the renormalized frequencies,
      ($\Omega_{\mathbf{q}\nu}$), the (black) dashed curve depicts the
      renormalized frequencies shifted by the real part of the self-energy,
      ($\Delta_{\mathbf{q}\nu}(\Omega_{\mathbf{q}\nu})$), and the (red) dotted
      curve shows the harmonic frequencies. The filled circle
      symbols show peak positions of the IXS spectra by our measurement. The
      filled diamond and square symbols depict the INS measurements by Raunio
      \textit{et al.}~\cite{Raunio-NaCl-1969} and by Schmunk and
      Winder,\cite{Schmunk-NaCl-1970} respectively. \label{fig:NaCl-sf} }
  \end{center}
\end{figure*}

Phonon measurements using INS were reported for KCl~\cite{Raunio-KCl-1969} and
NaCl~\cite{Raunio-NaCl-1969, Schmunk-NaCl-1970} at 300 K (room temperature).
Phonon frequencies measured in these studies are plotted in
Figs.~\ref{fig:KCl-sf} and \ref{fig:NaCl-sf}, for which the phonon frequency
values in Fig.~2 of Ref.~\onlinecite{Schmunk-NaCl-1970} were sampled using
WebPlotDigitizer.\cite{Rohatgi-2020} Recently, detailed calculation of phonon
band structure of NaCl was reported by Ravichandran and
Broido.\cite{Ravichandran-NaCl-calc-2018} They employed a phonon renormalization
approach to treat strong anharmonicity in NaCl, and showed good agreement of the
phonon frequencies with the measurement by Raunio {\it et
    al.}\cite{Raunio-NaCl-1969}

In this study, we investigate phonon spectral shapes of the LO modes of KCl and
NaCl using the IXS measurements and first-principles anahrmonic phonon
calculations. In particular, their LO-mode phonons near $\Gamma$-points are
shown to be strongly anharmonic. In Secs.~\ref{sec:method-of-experiment} and
\ref{sec:method-of-calculation}, methods of measurements and calculations are
described, respectively. In Sec.~\ref{sed:results-and-discussions}, first we
discuss about feasibility of the calculation results against the IXS
measurements by the peak positions and shapes of the phonon spectra, then we
analyze the calculated spectral functions in details.

\section{Method of experiment}
\label{sec:method-of-experiment}
The phonon spectra were measured using IXS at the BL35XU of the SPring-8
synchrotron.\cite{Baron-2000} The energy of 21.747 keV of the beam and Si
$\begin{pmatrix}11 & 11 & 11\end{pmatrix}$ backscattering setup with energy
resolution of $\sim$1.5 meV were used.

The phonon spectra of KCl and NaCl were measured around the $\Gamma$--L path of
$\mathbf{Q} = \begin{pmatrix}3-u & 3-u & 3-u\end{pmatrix}$ and $\mathbf{Q} =
  \begin{pmatrix} 3+u & 3+u & 3+u \end{pmatrix}$ with positive $u$,
respectively, where the points in the reciprocal spaces are represented with
respect to the reciprocal basis vectors of the conventional unit cell.
  { Their Q-resolutions were $\Delta \mathbf{Q} \sim
      \begin{pmatrix}0.03 & 0.02 & 0.04\end{pmatrix}$ and  $\Delta \mathbf{Q} \sim
      \begin{pmatrix}0.03 & 0.03 & 0.01\end{pmatrix}$}, respectively. For KCl,
those along the $\Gamma$--X path of $\mathbf{Q} = \begin{pmatrix}5-u & 1 &
               1\end{pmatrix}$ were also measured, and the Q-resolution was
  {$\Delta \mathbf{Q} \sim \begin{pmatrix}0.04 & 0.00 &
               0.04\end{pmatrix}$}. Due to the experimental setting, measured Q-points were
slightly deviated from the high-symmetry paths and {the
    coordinates are shown in Figs.~\ref{fig:KCl-G-L-spectra},
    \ref{fig:KCl-G-X-spectra}, \ref{fig:NaCl-G-L-spectra} along with the measured
    spectra}. Lorentzian functions were used to determine the peak positions of the
phonon spectra by least-squares fitting{, and the determined
    points are presented in Figs.~\ref{fig:KCl-sf} and \ref{fig:NaCl-sf}}.

\section{Method of calculation}
\label{sec:method-of-calculation}

\subsection{Phonon spectral function}

Anharmonic phonon calculations were performed in the stochastic self-consistent
harmonic approximation (SSCHA).\cite{Errea-SSCHA-2013, Errea-SSCHA-2014,
  Paulatto-SSCHA-2015, Bianco-SSCHA-2017, Bianco-SSCHA-2018, Ribeiro-SSCHA-2018,
  Unai-SSCHA-2019, Monacelli-SSCHA-2021} {There exist several
  software implementations of SSCHA. The SSCHA code~\cite{Monacelli-SSCHA-2021} is
  the software implementation of the original SSCHA method. The hiPhive
  code~\cite{hiPhive} has an implementation of a self-consistent harmonic
  approximation (SCHA) resembling Ref.~\onlinecite{Errea-SSCHA-2014}.} We employed
an iterative force-constants-fitting approach within the framework of SSCHA
which is considered equivalent to the methods reported in
Refs.~\onlinecite{van-Roekeghem-2020} and
\onlinecite{Hellman-2017-stocastic-sampling} {as implemented in
  the QSCAILD code and the TDEP code, respectively}.

SCHA force constants are given as
\begin{align}
  \label{eq:sc-fc2}
  \Phi_{l\kappa j,l'\kappa'j'} =
  \expval{\frac{\partial^2 V}{\partial u_{l\kappa j} \partial u_{l'\kappa' j'}}
  }_{\tilde{\rho}_{\Phi}},
\end{align}
where $V$ is the crystal potential and $u_{l\kappa j}$ is the atomic
displacement at the lattice point $l$, atom $\kappa$ of $l$, and Cartesian index
$j$. The angle bracket means average with respect to the density matrix of
harmonic phonons ${\tilde{\rho}_{\Phi}}$, where $\tilde{\rho}_{\Phi}$ is a
function of $\Phi$ at a temperature $T$, i.e., Eq.~(\ref{eq:sc-fc2}) is a
self-consistent equation at $T$. Displacements of atoms in direct space are
represented by superposition of phonon normal modes with amplitudes
$Q_{\mathbf{q}\nu}$, where $\mathbf{q}$ and $\nu$ are the wave vector and the
band index, respectively. The probability distribution function of each harmonic
phonon mode $(\mathbf{q}, \nu)$ at $T$, $P_{\mathbf{q}\nu}(T)$, is given
as~\cite{Barragan-Gil-sigma-2018,Thermal-Neutron-Scattering}
\begin{align}
  \label{eq:prob-dist}
   & P_{\mathbf{q}\nu}(T) = \frac{1}{\sqrt{2\pi
  \sigma^2_{\mathbf{q}\nu}(T)}} \exp\left[
    -\frac{1}{2}\frac{Q_{\mathbf{q}\nu}^2}{\sigma_{\mathbf{q}\nu}^2(T)}
  \right],                                      \\
   & \sigma_{\mathbf{q}\nu}^2(T) =
  \frac{\hbar}{2\Omega_{\mathbf{q}\nu}} [1 + 2n_{\mathbf{q}\nu}(T) ],
  \; n_{\mathbf{q}\nu}(T) = \frac{1}{e^{\hbar
        \Omega_{\mathbf{q}\nu}/k_\mathrm{B}T} - 1},\nonumber
\end{align}
where $\hbar$ and $k_\mathrm{B}$ denote the reduced Planck constant and the
Boltzmann constant, respectively. {$\Omega_{\mathbf{q}\nu}$ is
the phonon frequency as the solution of dynamical matrix of $\Phi$ in
Eq.~(\ref{eq:sc-fc2}). The density matrix ${\tilde{\rho}_{\Phi}}$ in
Eq.~(\ref{eq:sc-fc2}) corresponds to the product of $P_{\mathbf{q}\nu}(T)$,
i.e., $\prod_{\mathbf{q}\nu}P_{\mathbf{q}\nu}(T)$.}

The third-order force constants with respect to $\tilde{\rho}_{\Phi}$ are given
as~\cite{Bianco-SSCHA-2017, Ribeiro-SSCHA-2018, Unai-SSCHA-2019,
  Monacelli-SSCHA-2021}
\begin{align}
  \label{eq:sc-fc3}
  \Phi_{l\kappa j,l'\kappa'j',l''\kappa''j''}^{\tilde{\rho}_{\Phi}} =
  \expval{\frac{\partial^3
      V}{\partial u_{l\kappa j} \partial u_{l'\kappa' j'} \partial
      u_{l''\kappa'' j''}}
  }_{\tilde{\rho}_{\Phi}}.
\end{align}
With Eqs.~(\ref{eq:sc-fc2}) and (\ref{eq:sc-fc3}), spectral function of each
phonon mode~\cite{Bianco-SSCHA-2017} was calculated from the following
form:~\cite{Tadano-ALM-2018}
\begin{align}
  \label{eq:spectral-function}
  A_{\mathbf{q}\nu}(\omega) = \frac{1}{\pi} \frac{4\Omega^2_{\mathbf{q}\nu}
  \Gamma_{\mathbf{q}\nu}(\omega)}
  {\left[\omega^2 - \Omega^2_{\mathbf{q}\nu} -
  2\Omega_{\mathbf{q}\nu} \Delta_{\mathbf{q}\nu}(\omega) \right]^2
  + \left[ 2\Omega_{\mathbf{q}\nu}
    \Gamma_{\mathbf{q}\nu}(\omega) \right]^2},
\end{align}
where {$\omega$ is the phonon frequency sampled arbitrary in the
    calculation,} $\Delta_{\mathbf{q}\nu}(\omega)$ and
$\Gamma_{\mathbf{q}\nu}(\omega)$ denote the real and imaginary parts of the
self-energy of the bubble diagram, respectively, whose details are written in
Appendix~\ref{sec:self-energy}.

To compute Eqs.~(\ref{eq:sc-fc2}) and (\ref{eq:sc-fc3}), supercell approach was
used. Finite atomic displacements in supercells were generated by stochastically
sampling normal mode amplitudes $Q_{\mathbf{q}\nu}$ on the probability
distribution functions of $P_{\mathbf{q}\nu}(T)$ at the commensurate q-points.
Force constants were obtained from atomic displacements and forces by linear
regression,\cite{Tadano-ALM-2018} where the forces were calculated using
first-principles calculation. More computational details are given in the next
section.

\subsection{Computational details}
\label{sec:computational-details}

For the conventional unit cell models, experimental lattice parameters of 6.29
and 5.64 {\AA} at 300 K for KCl and NaCl~\cite{Barrett-NaCl-KCl-1954},
respectively, were used. For the supercell phonon calculation, we employed the
phonopy~\cite{phonopy} and phono3py~\cite{phono3py} codes. Non-analytical term
correction~\cite{Pick-1970,Gonze-1994,Gonze-1997} was applied to dynamical
matrices to treat long range dipole-dipole interactions. For force constants
fitting, the ALM code~\cite{Tadano-ALM-2018} was used. Supercells of $2\times
  2\times 2$ expansion of conventional unit cells of KCl and NaCl were used for
most of the harmonic and anharmonic phonon calculations. In addition, $4\times
  4\times 4$ supercells were used for the calculations of the harmonic force
constants to replace harmonic part of the SSCHA force constants calculated with
the $2\times 2\times 2$ supercells, i.e., we approximate $\Phi$ of
Eq.~(\ref{eq:sc-fc2}) by $\Phi_{4\times 4\times 4} \sim \Phi_{4\times 4\times
    4}^{(0)} + \Phi_{2\times 2\times 2} - \Phi_{2\times 2\times 2}^{(0)}$, where
$\Phi^{(0)}$ denotes the harmonic force constants and the subscript indicates
the supercell size.

For the first-principles calculations, we employed the plane-wave basis
projector augmented wave (PAW) method~\cite{PAW-Blochl-1994} within the
framework of DFT as implemented in the VASP
code.\cite{VASP-Kresse-1995,VASP-Kresse-1996,VASP-Kresse-1999} The generalized
gradient approximation (GGA) of Perdew, Burke, and Ernzerhof revised for solids
(PBEsol)~\cite{PBEsol} was used as the exchange correlation potential. For PAW
datasets of atoms, 3p electrons for K and 2p electrons for Na were treated as
valence. Static dielectric constants and Born effective charges were calculated
with primitive cells from density functional perturbation theory (DFPT) as
implemented in the VASP code\cite{Gajdos-2006,Wu-2005}. A plane-wave energy
cutoff of 500 eV was employed for the supercell force calculations and 750 eV
for the DFPT calculations. Reciprocal spaces were sampled by half-shifted
$4\times 4\times 4$ meshes for the $2\times 2\times 2$ supercells, half-shifted
$2\times 2\times 2$ meshes for the $4\times 4\times 4$ supercells, and the
$\Gamma$-centered $8\times 8 \times 8$ meshes for the primitive cells. For each
harmonic force constants calculation, atoms in 20 supercells were randomly
displaced in directions with a fixed distance of 0.03 \AA~from their equilibrium
positions. The high-frequency dielectric constants ($\epsilon_\infty$) of KCl
and NaCl were obtained as 2.365 and 2.546, and the Born effective charges as
$\pm 1.129$ and $\pm 1.096$, respectively.

The SSCHA force constants were obtained by iterating phonon calculations.
Initial phonon calculation was performed with small displacements (0.03 \AA) in
supercells. At every iteration step, 20 supercells with random atomic
displacements as a batch were generated as given by Eq.~(\ref{eq:prob-dist}) at
300 K using force constants that were calculated with supercell
displacement-force datasets of up to previous 50 batches (1000 supercells).
Then, forces of the supercells with the generated displacements were calculated
using the VASP code. This process was repeated 100 times, and we took the last
force constants as the converged SSCHA force constants. {Details
    about the convergence is summarized in Appendix \ref{sec:convergence-sscha}.}
For the calculation of Eq.~(\ref{eq:sc-fc3}), 4000 supercells with random atomic
displacements generated from the SSCHA force constants according to
Eq.~(\ref{eq:prob-dist}) were used for the fitting of third-order force
constants. Commensurate q-points were sampled for Eq.~(\ref{eq:prob-dist}),
which guarantees the generation of real valued displacements in the supercells.
To perform systematic calculations presented above, we employed the AiiDA
environment~\cite{AiiDA} with the AiiDA-VASP~\cite{AiiDA-VASP} and
AiiDA-phonoxpy~\cite{AiiDA-phonoxpy} plugins. {For the
    calculations of the spectral functions, self-energies, and weighted joint
    density of states, the q-points were sampled on regular grids of $300 \times 300
      \times 300$ mesh, and the phonon frequencies were uniformly sampled at 2001
    points from 0 meV to about twice the highest renormalied phonon frequencies.}

\section{Results and discussions}
\label{sed:results-and-discussions}

\subsection{Phonon band structures}

Phonon structures of KCl and NaCl are presented in Figs.~\ref{fig:KCl-sf} and
\ref{fig:NaCl-sf}, respectively. The left panels show the
calculated spectral functions. The points and curves in the right panels depict
experimentally measured and calculated results, respectively. The LO-mode
frequencies that we measured agree well with the INS measurements reported
in Refs.~\onlinecite{Raunio-KCl-1969, Raunio-NaCl-1969, Schmunk-NaCl-1970} for
KCl and NaCl.

The calculated spectral functions of the LO modes near the $\Gamma$-points show
side bands due to their strong anharmonicity. Except for the LO modes, the
spectra show clear peaks. We can see the peak positions underestimate the
experiments systematically. The underestimation is largest for the LO modes
near the $\Gamma$-points. However we are satisfied with the current
level of the agreements between the calculations and experiments since it is
expected that this level of the errors hardly affect the shapes of the
calculated spectral functions.

The three curves in the right panel in each of Figs.~\ref{fig:KCl-sf} and
\ref{fig:NaCl-sf} show harmonic frequencies, $\Omega_{\mathbf{q}\nu}^{(0)}$,
calculated from $\Phi^{(0)}$, renormalized harmonic frequencies,
$\Omega_{\mathbf{q}\nu}$, obtained from $\Phi$ of Eq.~(\ref{eq:sc-fc2}), and
$\Omega_{\mathbf{q}\nu}$ shifted by real parts of the self-energies at
$\Omega_{\mathbf{q}\nu}$, $\Omega_{\mathbf{q}\nu} +
  \Delta_{\mathbf{q}\nu}(\Omega_{\mathbf{q}\nu})$. By cancellation between the
renormalizations $\Omega_{\mathbf{q}\nu} - \Omega_{\mathbf{q}\nu}^{(0)}$ and the
shifts $\Delta_{\mathbf{q}\nu}(\Omega_{\mathbf{q}\nu})$, the harmonic
frequencies $\Omega_{\mathbf{q}\nu}^{(0)}$ become close to
$\Omega_{\mathbf{q}\nu} + \Delta_{\mathbf{q}\nu}(\Omega_{\mathbf{q}\nu})$ and
show even better agreements with the experiments than $\Omega_{\mathbf{q}\nu} +
  \Delta_{\mathbf{q}\nu}(\Omega_{\mathbf{q}\nu})$ for the LO modes. However, we
consider this is a specific result for KCl and NaCl and not general tendency.

As shown in the left panels of Figs.~\ref{fig:KCl-sf} and
\ref{fig:NaCl-sf},
for the sharp spectral functions, $\Omega_{\mathbf{q\nu}} +
  \Delta_{\mathbf{q}\nu}(\Omega_{\mathbf{q}\nu})$ are expected to agree well with
their peak positions. Near the $\Gamma$-point, since the LO modes exhibit the
broad spectral functions, $\Omega_{\mathbf{q},\text{LO}} +
  \Delta_{\mathbf{q},\text{LO}}(\Omega_{\mathbf{q},\text{LO}})$ are unable to
represent their peak positions. This is an effect of the strong anharmonicity.

\subsection{Comparisons of spectral shapes}

\begin{figure*}[ht]
  \begin{center}
    \includegraphics[width=0.95\linewidth]{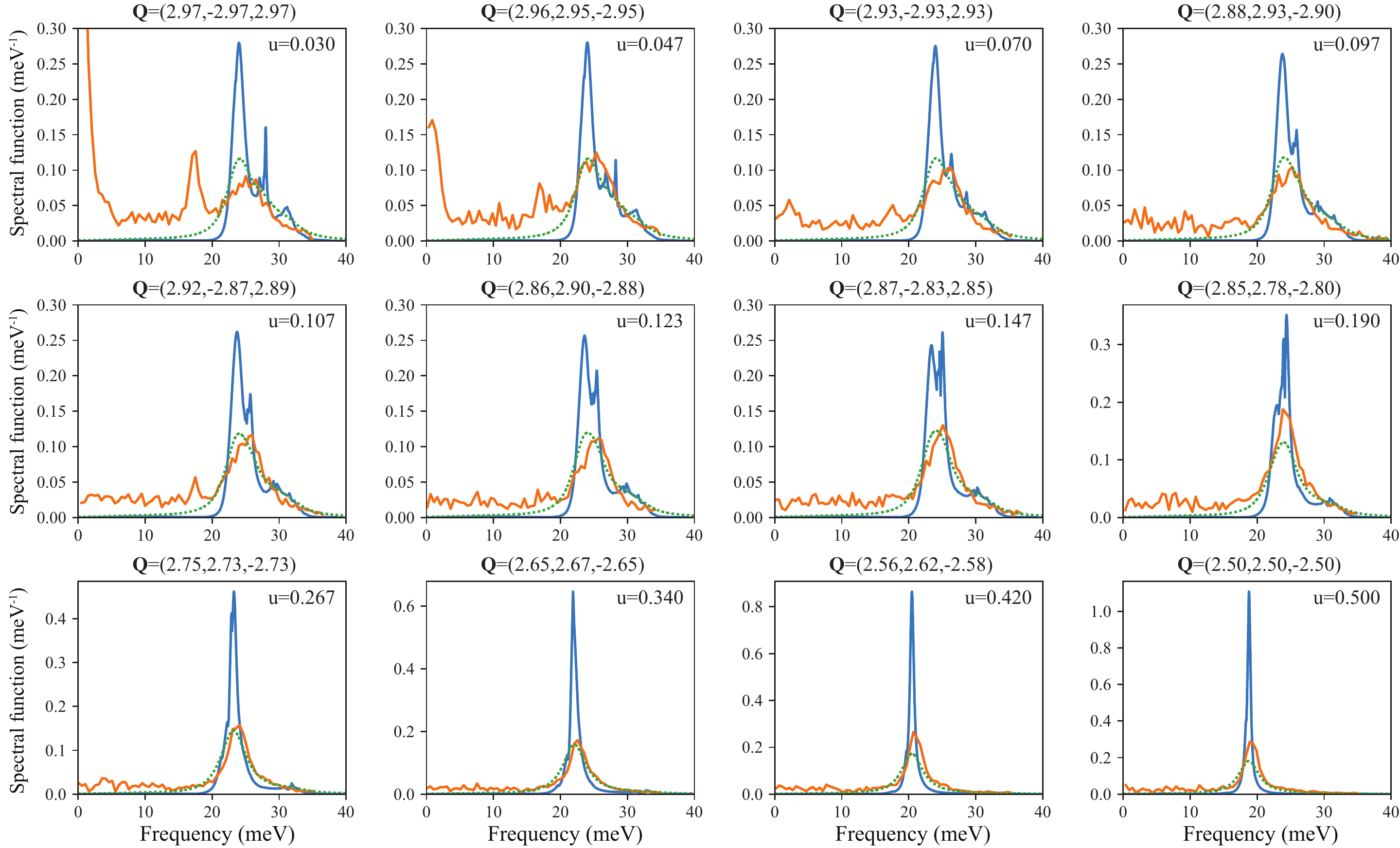}
    \caption{KCl spectra measured by IXS at Q-points (see Fig.~\ref{fig:KCl-sf})
      and calculated phonon spectral functions of the LO mode at q-points near
      the $\Gamma$--L path. The (orange) solid curves with broader
      peaks are the IXS measurements. The (blue) solid and (green) dotted curves
      show the calculated spectral functions
      ($A_{\mathbf{q},\text{LO}}(\omega)$) and those smeared by the Lorentzian
      function with the 1.5 meV scale parameter, respectively. The q-points of
      the calculation were chosen from the grid points on the $\Gamma$--L path at
      which their distances from the $\Gamma$-point are closest to those of the
      measured Q-points. The q-point is represented by $\mathbf{q}=(u, u, u)$
      with respect to the conventional basis of the Bilbao crystallographic
      server,\cite{Aroyo-2014} where $u=0.5$ gives the L-point.
      \label{fig:KCl-G-L-spectra} }
  \end{center}
\end{figure*}

\begin{figure*}[ht]
  \begin{center}
    \includegraphics[width=0.95\linewidth]{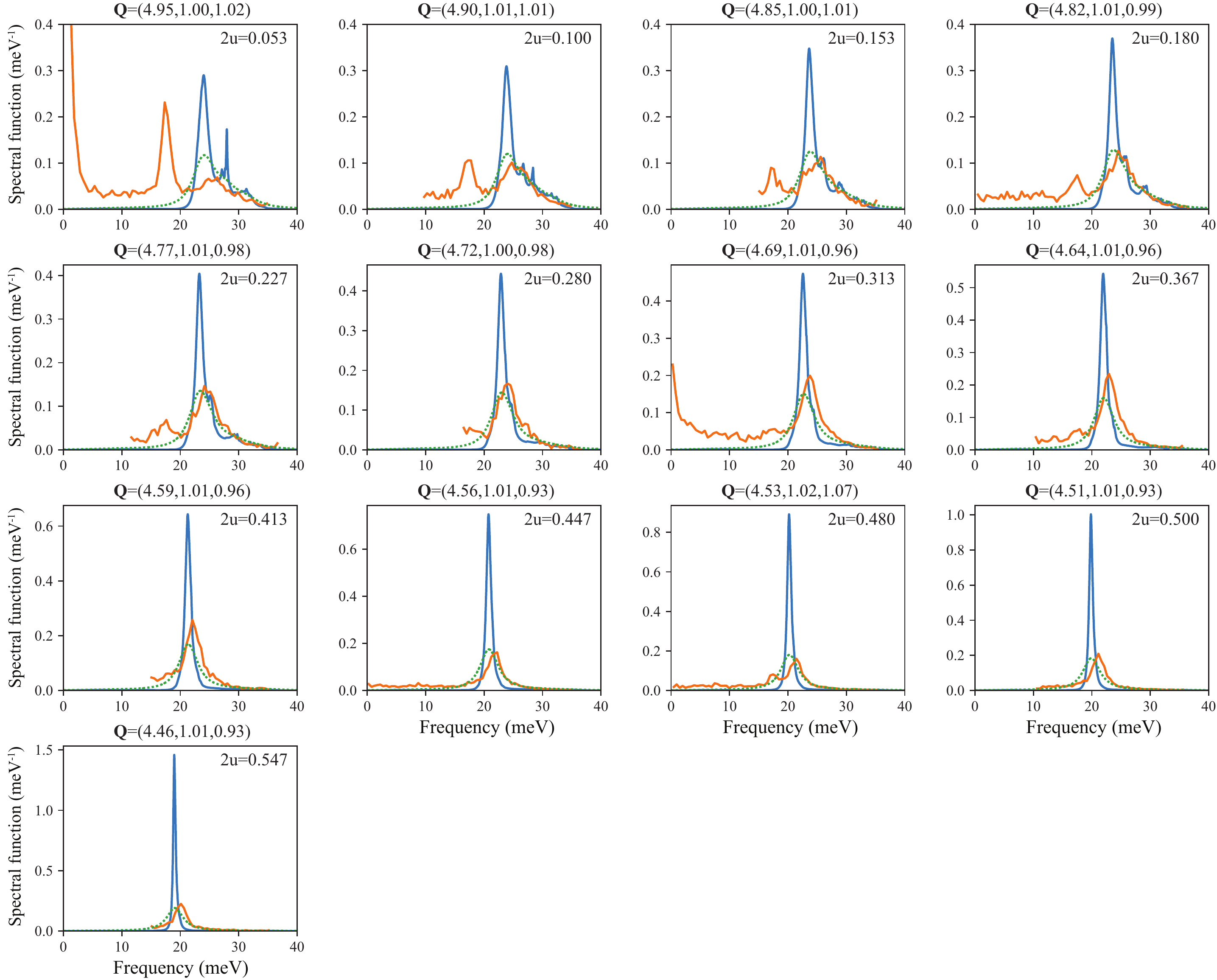}
    \caption{KCl spectra measured by IXS at Q-points (see Fig.~\ref{fig:KCl-sf})
      and calculated phonon spectral functions of the LO mode at q-points near
      the $\Gamma$--X path. The (orange) solid curves with broader peaks are the
      IXS measurements. The (blue) solid and (green) dotted curves show the
      calculated spectral functions ($A_{\mathbf{q},\text{LO}}(\omega)$) and
      those smeared by the Lorentzian function with the 1.5 meV scale parameter,
      respectively. The q-points of the calculation were chosen from the grid
      points on the $\Gamma$--X path at which their distances from the
      $\Gamma$-point are closest to those of the measured Q-points. The q-point
      is represented by $\mathbf{q}=(2u, 0, 0)$ with respect to the conventional
      basis of the Bilbao crystallographic server,\cite{Aroyo-2014} where $2u=1$
      gives the X-point.
      \label{fig:KCl-G-X-spectra}
    }
  \end{center}
\end{figure*}

\begin{figure*}[ht]
  \begin{center}
    \includegraphics[width=0.95\linewidth]{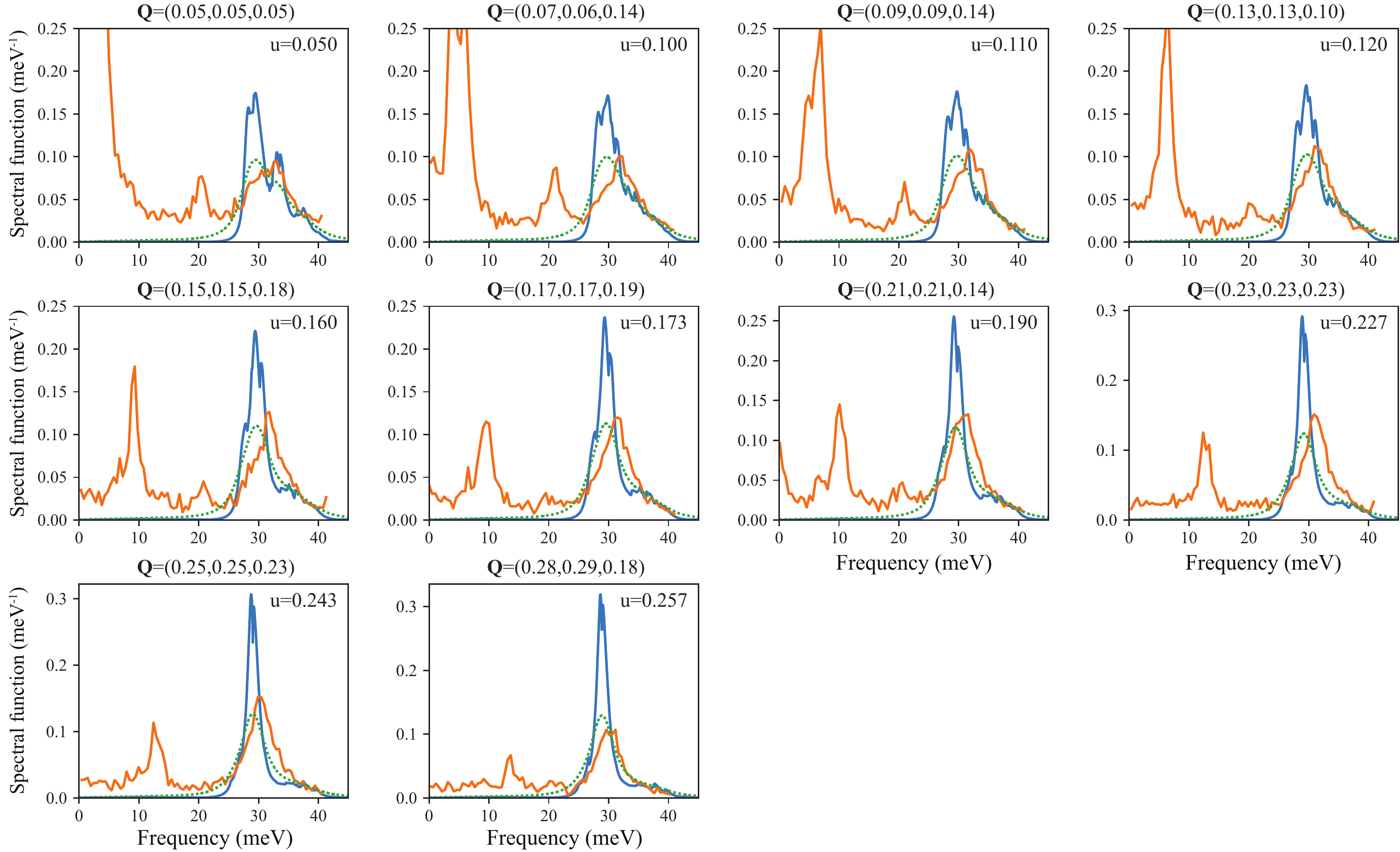}
    \caption{NaCl spectra measured by IXS at Q-points (see
      Fig.~\ref{fig:NaCl-sf}) and calculated phonon spectral functions of the LO
      mode at q-points near the $\Gamma$--L path. The (orange) solid curves with
      broader peaks are the IXS measurements. The (blue) solid and (green)
      dotted curves show the calculated spectral functions
      ($A_{\mathbf{q},\text{LO}}(\omega)$) and those smeared by the Lorentzian
      function with the 1.5 meV scale parameter, respectively. The q-points of
      the calculation were chosen from the grid points on the $\Gamma$--L path at
      which their distances from the $\Gamma$-point are closest to those of the
      measured Q-points. The q-point is represented by $\mathbf{q}=(u, u, u)$
      with respect to the conventional basis of the Bilbao crystallographic
      server,\cite{Aroyo-2014} where $u=0.5$ gives the L-point.
      \label{fig:NaCl-G-L-spectra} }
  \end{center}
\end{figure*}

{The measured IXS spectra and calculated spectral functions of
the LO modes are compared in Figs.~\ref{fig:KCl-G-L-spectra},
\ref{fig:KCl-G-X-spectra}, and \ref{fig:NaCl-G-L-spectra}. The calculated
results show the LO-mode contributions only, i.e., what are presented are
$A_{\mathbf{q},\text{LO}}(\omega)$. Finite Q-resolution effects make acoustic
and/or optical branches with different polarization observed near the
$\Gamma$-points in the IXS experiments. The LO-mode spectra show the broader
peaks near the $\Gamma$-points. With increasing distance from the
$\Gamma$-points, the spectral peaks become shaper. }

{The IXS spectra show broader spectral shapes than the calculated
spectral functions due to the finite IXS energy resolution. To include this
effect in the calculations, the spectral functions were smeared by the
Lorentzian function with a scale parameter of the 1.5 meV. The smeared
spectral functions are also shown in Figs.~\ref{fig:KCl-G-L-spectra},
\ref{fig:KCl-G-X-spectra}, and \ref{fig:NaCl-G-L-spectra}. We can see that
general trend of the IXS spectral shapes are well reproduced by them.
Therefore, we consider that details of the anharmonic spectra can be discussed
from the calculated spectral functions.}

\subsection{Spectral functions and self-energies}

\begin{figure*}[ht]
  \begin{center}
    \includegraphics[width=0.95\linewidth]{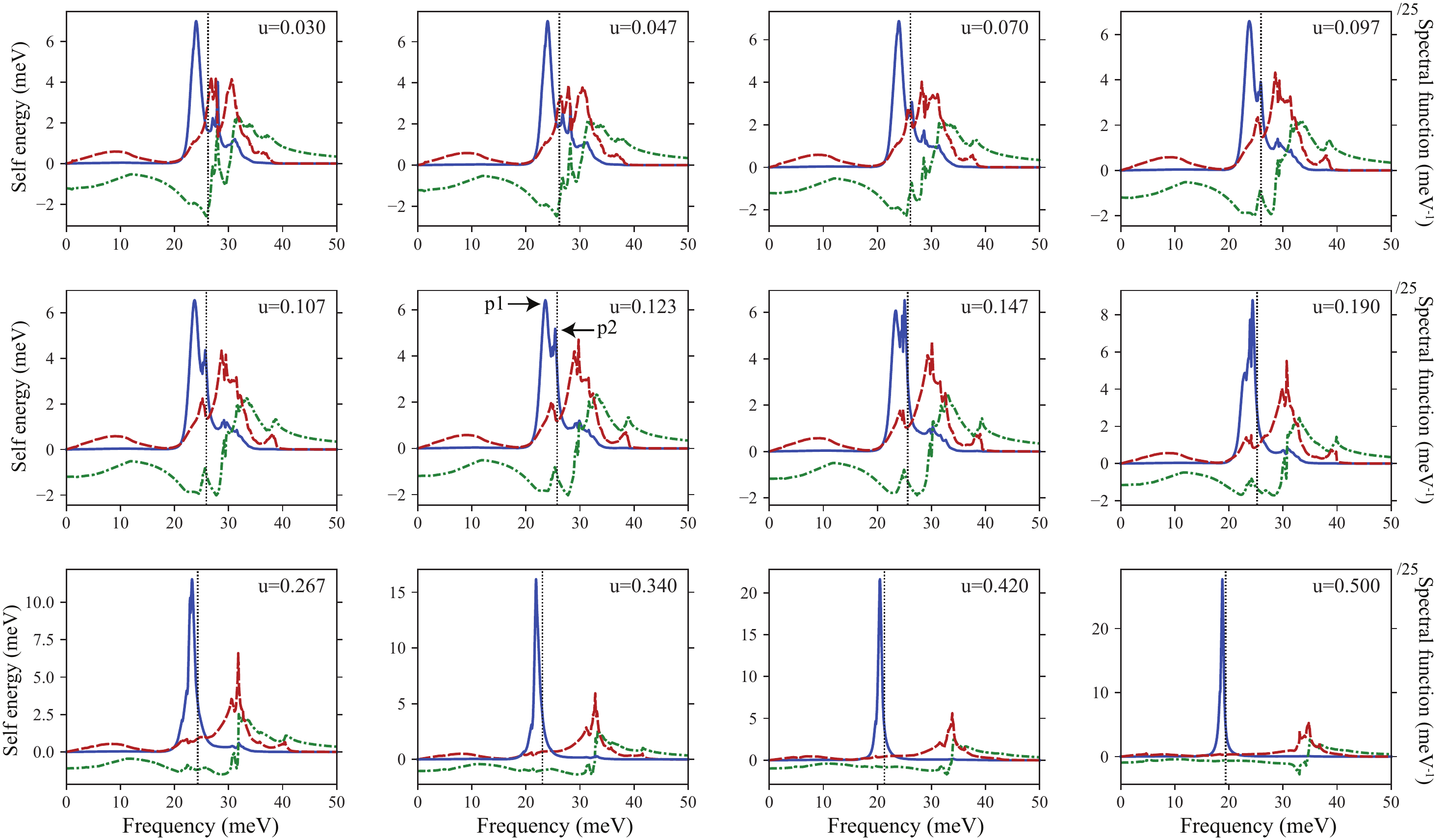}
    \caption{Real and imaginary parts of self-energies and renormalized
      frequencies of KCl at q-points on the $\Gamma$--L path. Each panel
      corresponds to that in Fig.~\ref{fig:KCl-G-L-spectra}. The dashed-dotted
      (green) and dashed (red) curves show the real
      ($\Delta_{\mathbf{q},\text{LO}}(\omega)$) and imaginary
      ($\Gamma_{\mathbf{q},\text{LO}}(\omega)$) parts of the self-energies,
      respectively. The solid (blue) curves show the spectral functions
      ($A_{\mathbf{q},\text{LO}}(\omega)$), that are the same as those in
      Fig.~\ref{fig:KCl-G-L-spectra}. The vertical dotted lines indicate the
      renormalized frequencies ($\Omega_{\mathbf{q},\text{LO}}$). The small
      arrows in the panel $u$=0.123 depict two main peaks that change their
      intensity ratio at q-points along the $\Gamma$--L path.
      \label{fig:KCl-G-L-self-energy}}
  \end{center}
\end{figure*}

\begin{figure*}[ht]
  \begin{center}
    \includegraphics[width=0.95\linewidth]{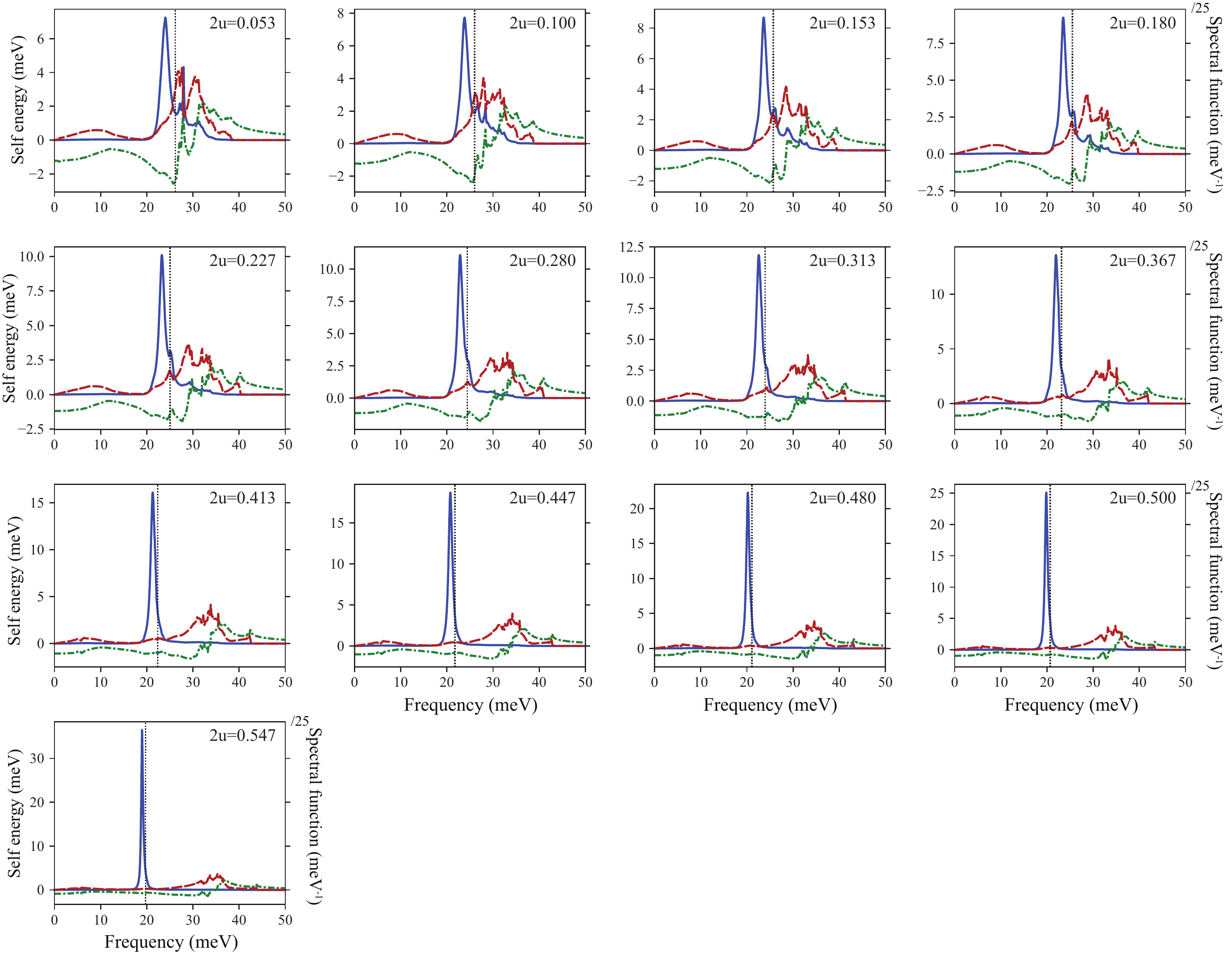}
    \caption{Real and imaginary parts of self-energies and renormalized
      frequencies of KCl at q-points on the $\Gamma$--X path. Each panel
      corresponds to that in Fig.~\ref{fig:KCl-G-X-spectra}. The dashed-dotted
      (green) and dashed (red) curves show the real
      ($\Delta_{\mathbf{q},\text{LO}}(\omega)$) and imaginary
      ($\Gamma_{\mathbf{q},\text{LO}}(\omega)$) parts of the self-energies,
      respectively. The solid (blue) curves show the spectral functions
      ($A_{\mathbf{q},\text{LO}}(\omega)$), that are the same as those in
      Fig.~\ref{fig:KCl-G-X-spectra}. The vertical dotted lines indicate the
      renormalized frequencies ($\Omega_{\mathbf{q},\text{LO}}$).
      \label{fig:KCl-G-X-self-energy}}
  \end{center}
\end{figure*}

\begin{figure*}[ht]
  \begin{center}
    \includegraphics[width=0.95\linewidth]{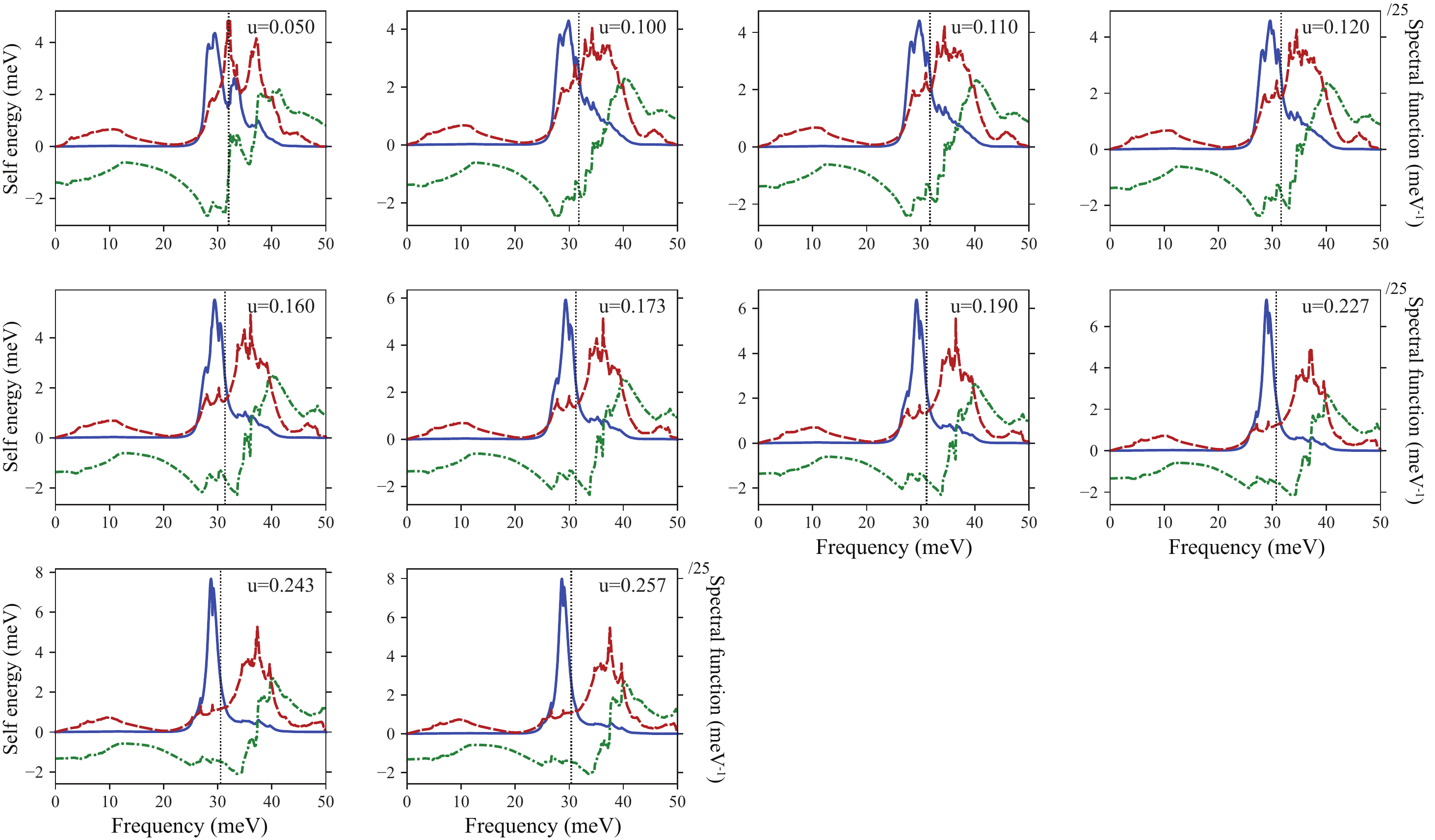}
    \caption{Real and imaginary parts of self-energies and renormalized
      frequencies of NaCl at q-points on the $\Gamma$--L path. Each panel
      corresponds to that in Fig.~\ref{fig:NaCl-G-L-spectra}. The solid (blue)
      curves show the spectral functions ($A_{\mathbf{q},\text{LO}}(\omega)$).
      The dashed-dotted (green) and dashed (red) curves show the real
      ($\Delta_{\mathbf{q},\text{LO}}(\omega)$) and imaginary
      ($\Gamma_{\mathbf{q},\text{LO}}(\omega)$) parts of the self-energies,
      respectively. The solid (blue) curves show the spectral functions, that
      are the same as those in Fig.~\ref{fig:NaCl-G-L-spectra}. The vertical
      dotted lines indicate the renormalized frequencies
      ($\Omega_{\mathbf{q},\text{LO}}$). \label{fig:NaCl-G-L-self-energy}}
  \end{center}
\end{figure*}

\begin{figure*}[ht]
  \begin{center}
    \includegraphics[width=0.95\linewidth]{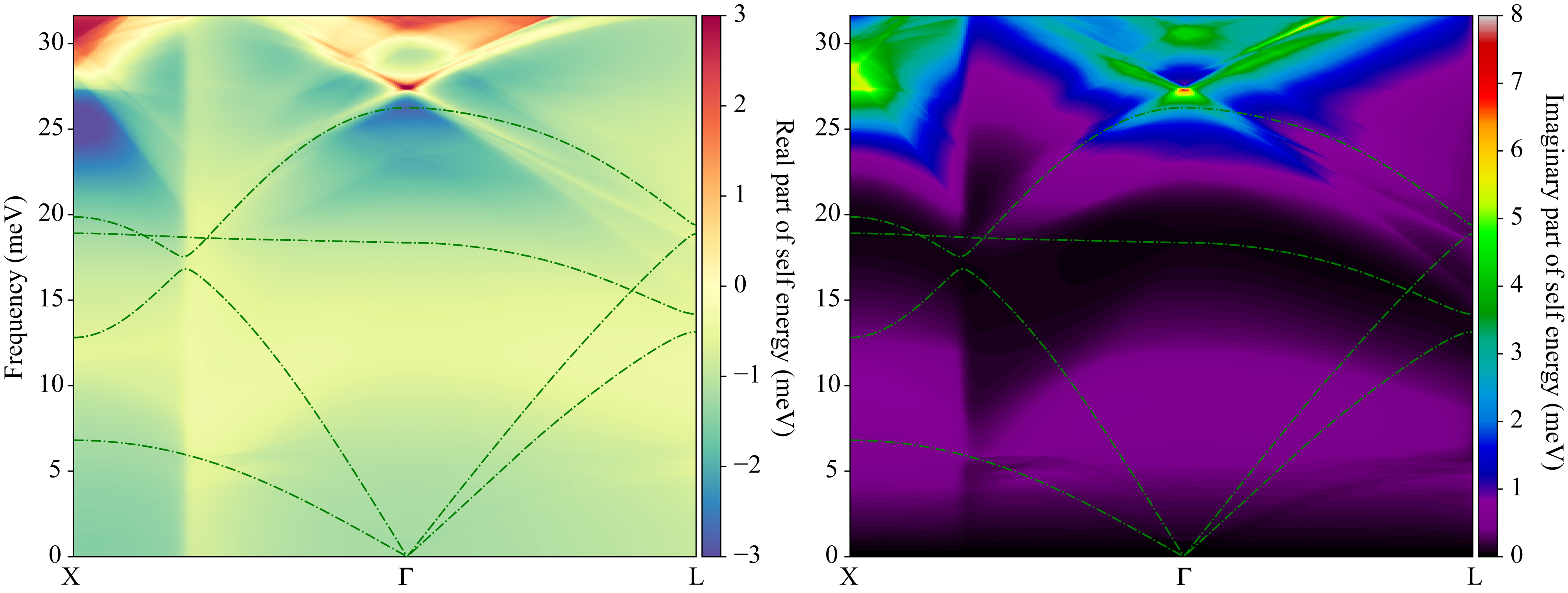}
    \caption{(Left panel) Real ($\Delta_{\mathbf{q},\text{LO}}(\omega)$) and
      (right panel) imaginary ($\Gamma_{\mathbf{q},\text{LO}}(\omega)$) parts of
      self-energies obtained for the LO mode of KCl at 300 K. The (green)
      dashed-dotted curve show the renormalized frequencies,
      $\Omega_{\mathbf{q}\nu}$. \label{fig:KCl-self-energy-heatmap}}
  \end{center}
\end{figure*}

\begin{figure*}[ht]
  \begin{center}
    \includegraphics[width=0.95\linewidth]{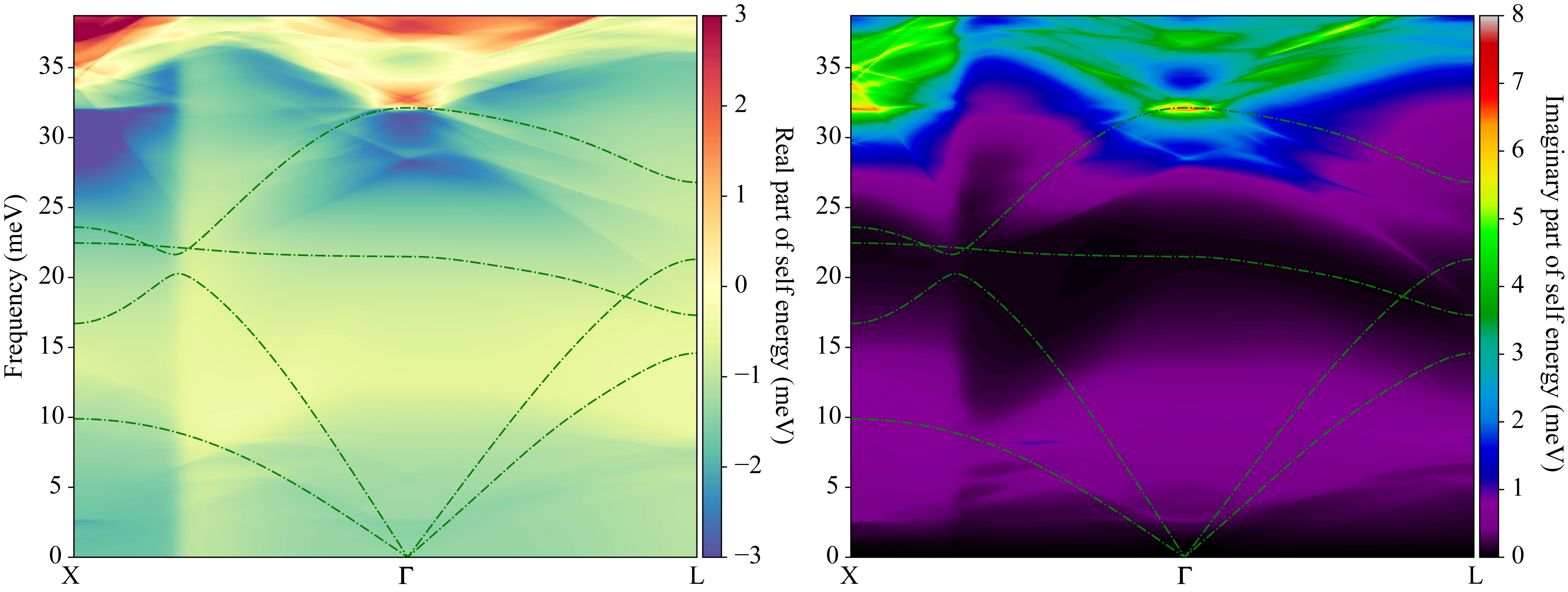}
    \caption{(Left panel) Real ($\Delta_{\mathbf{q},\text{LO}}(\omega)$) and
      (right panel) imaginary ($\Gamma_{\mathbf{q},\text{LO}}(\omega)$) parts of
      self-energies obtained for the LO mode of NaCl at 300 K. The (green)
      dashed-dotted curve show the renormalized frequencies,
      $\Omega_{\mathbf{q}\nu}$. \label{fig:NaCl-self-energy-heatmap}}
  \end{center}
\end{figure*}

The spectral function $A_{\mathbf{q}\nu}(\omega)$ in
Eq.~(\ref{eq:spectral-function}) is obtained from the renormalized frequency
$\Omega_{\mathbf{q}\nu}$ and the real and imaginary parts of the self-energy,
$\Delta_{\mathbf{q}\nu}(\omega)$ and $\Gamma_{\mathbf{q}\nu}(\omega)$,
respectively. In this section, the spectral functions of the LO modes,
$A_{\mathbf{q},\text{LO}}(\omega)$, are analyzed using
$\Omega_{\mathbf{q},\text{LO}}$, $\Delta_{\mathbf{q},\text{LO}}(\omega)$, and
$\Gamma_{\mathbf{q},\text{LO}}(\omega)$ in detail.

  {$A_{\mathbf{q},\text{LO}}(\omega)$, $\Delta_{\mathbf{q},\text{LO}}(\omega)$,
    $\Gamma_{\mathbf{q},\text{LO}}(\omega)$, and $\Omega_{\mathbf{q},\text{LO}}$}
at the same q-points used in
Figs.~\ref{fig:KCl-G-L-spectra}, \ref{fig:KCl-G-X-spectra}, and
\ref{fig:NaCl-G-L-spectra}
are shown in Figs.~\ref{fig:KCl-G-L-self-energy}, \ref{fig:KCl-G-X-self-energy},
and \ref{fig:NaCl-G-L-self-energy}, respectively. As
Eq.~(\ref{eq:spectral-function}) indicates, the spectral functions are expected
to have strong peaks around $\omega \sim \Omega_{\mathbf{q}\nu} +
  \Delta_{\mathbf{q}\nu}(\omega)$, which we call main peaks. {In
    these figures, the main peaks are located a few meV below
    $\Omega_{\mathbf{q},\text{LO}}$ since $\Delta_{\mathbf{q},\text{LO}}(\omega)$
    have negative values near the main peaks.}

The spectral functions show other multiple peaks than their main peaks. {These
    peaks become stronger when $\Gamma_{\mathbf{q},\text{LO}}(\omega)$ is larger
    at $\omega$ close to $\Omega_{\mathbf{q},\text{LO}}$. This condition is
    satisfied at the q-points near the $\Gamma$-points. The large
    $\Gamma_{\mathbf{q},\text{LO}}(\omega)$ is mainly the result of energy and
    momentum conversations of three phonon scatterings of the class formally
    denoted as $(\mathbf{q}, \omega) \rightarrow (\mathbf{q}', \omega_{\nu'}) +
      (\mathbf{q}'', \omega_{\nu''})$, which is discussed in
    Sec.~\ref{sec:three-phonon-scattering}.}

  {The distributions of $\Delta_{\mathbf{q},\text{LO}}(\omega)$ and
    $\Gamma_{\mathbf{q},\text{LO}}(\omega)$ are presented in
    Figs.~\ref{fig:KCl-self-energy-heatmap} and
    \ref{fig:NaCl-self-energy-heatmap} similarly to the left panels of
    Figs.~\ref{fig:KCl-sf} and \ref{fig:NaCl-sf}, respectively, where
    $\Omega_{\mathbf{q}\nu}$ are superimposed on these figures instead of
    $\Omega_{\mathbf{q}\nu} + \Delta_{\mathbf{q}\nu}(\Omega_{\mathbf{q}\nu})$.
    The overall distributions are similar between KCl and NaCl except for their
    frequency scales. Each of them is roughly symmetric between the $\Gamma$--L
    and $\Gamma$--X path sides. In the low frequency domains, the distributions
    are relatively flat mainly because the term $(n_{\mathbf{q}'\nu'} -
      n_{\mathbf{q}''\nu''})$ in Eq.~(\ref{eq:imag-self-energy}) is well cancelled
    as discussed in Sec.~\ref{sec:three-phonon-scattering}. We can see the
    characteristic distributions in the vicinities of the highest frequencies of
    $\Omega_{\mathbf{q},\text{LO}}$ near the $\Gamma$ points, which provides
    interesting spectral shapes. Large $\Gamma_{\mathbf{q},\text{LO}}(\omega)$
    are also found at the high frequency domains near the X-points, however
    $A_{\mathbf{q},\text{LO}}(\omega)$ are less anharmonic since
    $\Omega_{\mathbf{q},\text{LO}}$ are low enough to avoid passing through
    these $(\mathbf{q},\omega)$ domains.}

\subsection{Three phonon scattering processes}
\label{sec:three-phonon-scattering}

\begin{figure*}[ht]
  \begin{center}
    \includegraphics[width=0.95\linewidth]{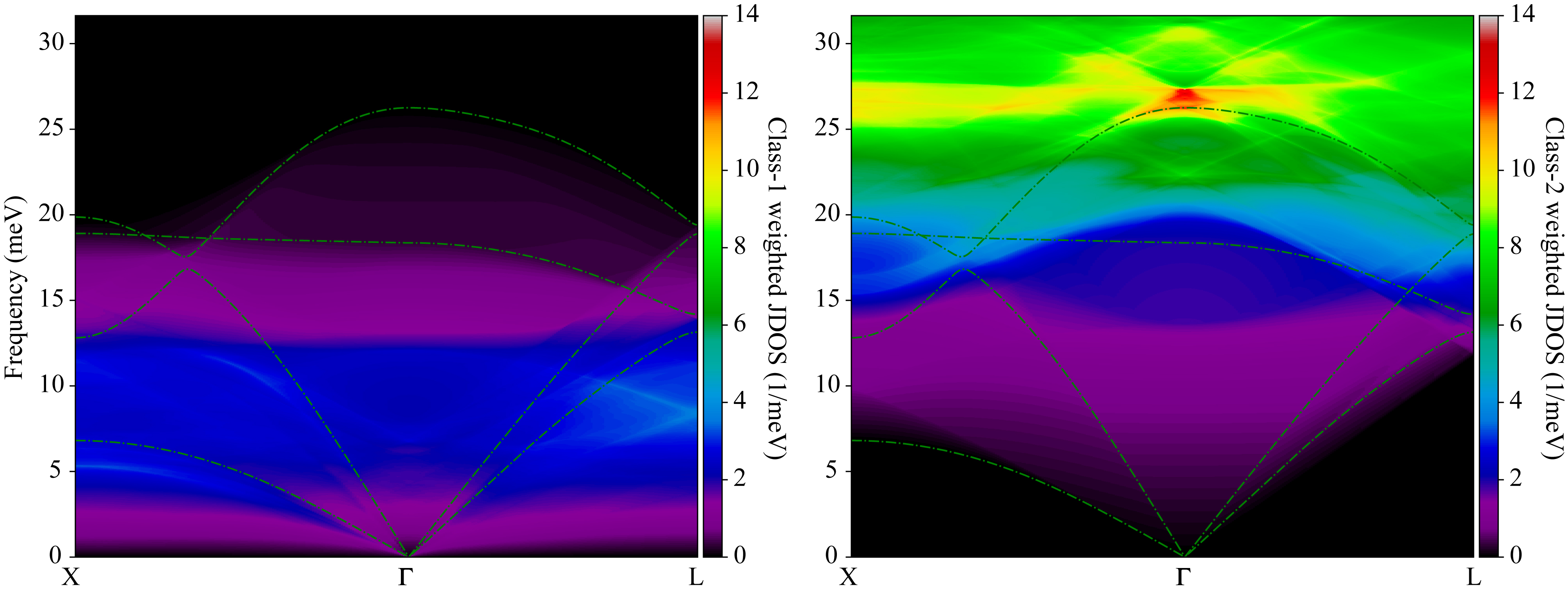}
    \caption{Weighted JDOS of class 1 (left panel, $N_2^{(1)}(\mathbf{q},
      \omega)$) and class 2 (right panel, $N_2^{(2)}(\mathbf{q}, \omega)$) of KCl
    at 300 K. The (green) dashed-dotted curve show the renormalized frequencies,
    $\Omega_{\lambda}$. \label{fig:KCl-jdos-heatmap}}
  \end{center}
\end{figure*}

\begin{figure*}[ht]
  \begin{center}
    \includegraphics[width=0.95\linewidth]{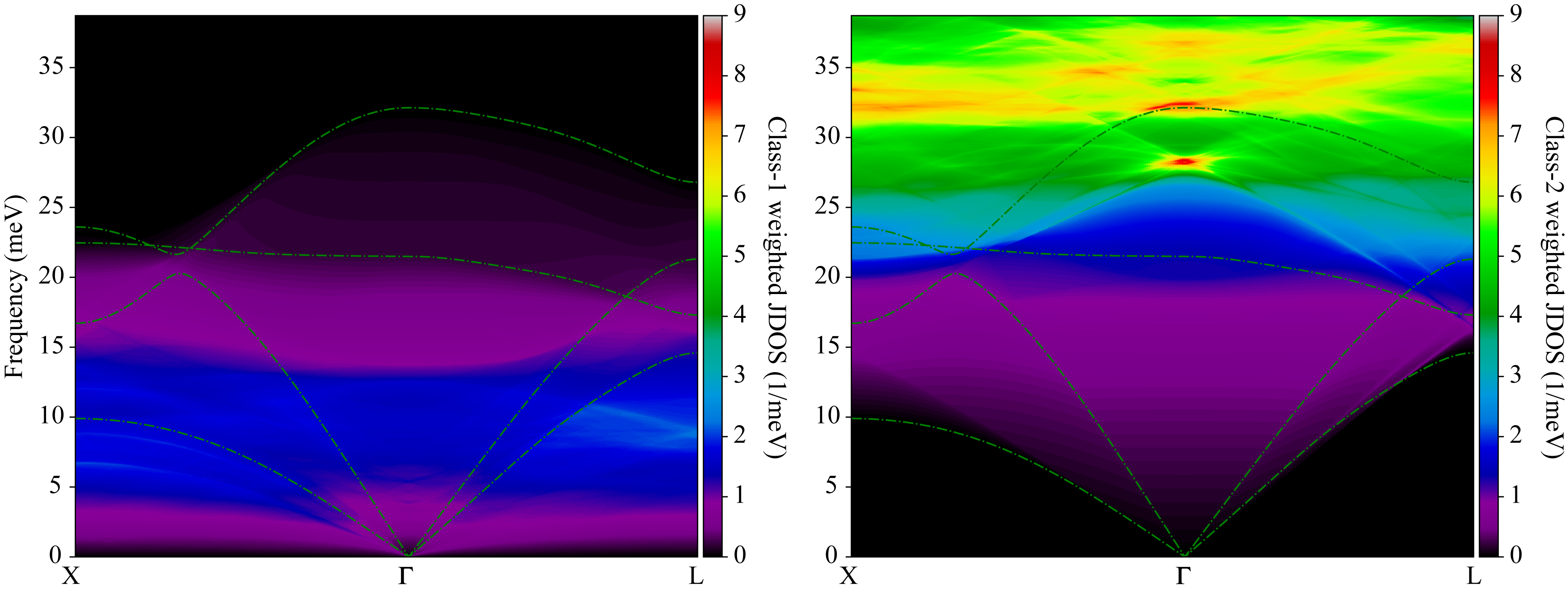}
    \caption{Weighted JDOS of class 1 (left panel, $N_2^{(1)}(\mathbf{q},
      \omega)$) and class 2 (right panel, $N_2^{(2)}(\mathbf{q}, \omega)$) of
    NaCl at 300 K. The (green) dashed-dotted curve show the renormalized
    frequencies, $\Omega_{\lambda}$. \label{fig:NaCl-jdos-heatmap}}
  \end{center}
\end{figure*}

The imaginary part of self-energy $\Gamma_\lambda(\omega)$ in
Eq.~(\ref{eq:imag-self-energy}) is determined by detailed combinations of
$\Phi_{\lambda\lambda'\lambda''}^{\tilde{\rho}_{\Phi}}$ and energy conservations
weighted by phonon occupation numbers. Due to translational symmetry,
$\Phi_{\lambda\lambda'\lambda''}^{\tilde{\rho}_{\Phi}}$ contains momentum
conservation as represented by $\Delta(\mathbf{q}+\mathbf{q}'+\mathbf{q}'')$ in
Eq.~(\ref{eq:ph-ph-strength}). Like the previous
works~\cite{phono3py,Mizokami-2018}, in this section, we discuss impacts of the
energy and momentum conservations of three phonon scatterings by introducing
weighted joint-density-of-states (JDOS). The weighted JDOS $N_2(\mathbf{q},
  \omega)$ is defined by replacing $\frac{18 \pi}{\hbar^2}\bigl|
  \Phi_{-\lambda\lambda'\lambda''}^{\tilde{\rho}_{\Phi}} \bigl|^2$ in
Eq.~(\ref{eq:imag-self-energy}) by
$\frac{1}{N}\Delta(-\mathbf{q}+\mathbf{q}'+\mathbf{q}'')$:
\begin{align}
  \label{eq:wjdos-sum}
  N_2(\mathbf{q}, \omega) = &
  N_2^{(1)}(\mathbf{q}, \omega) + N_2^{(2)}(\mathbf{q}, \omega),
\end{align}
where
\begin{align}
  \label{eq:wjdos-n1}
  N_2^{(1)}(\mathbf{q}, \omega) = &
  \frac{1}{N} \sum_{\lambda'\lambda''}
  \Delta(-\mathbf{q}+\mathbf{q}'+\mathbf{q}'') (n_{\lambda'} - n_{\lambda''})
  \nonumber
  \\
                                  &
  \times [ \delta( \omega + \Omega_{\lambda'} - \Omega_{\lambda''})
    - \delta(  \omega - \Omega_{\lambda'} + \Omega_{\lambda''})],
  \\
  \label{eq:wjdos-n2}
  N_2^{(2)}(\mathbf{q}, \omega) = &
  \frac{1}{N} \sum_{\lambda'\lambda''}
  \Delta(-\mathbf{q}+\mathbf{q}'+\mathbf{q}'') (n_{\lambda'}+ n_{\lambda''}+1)
  \nonumber
  \\
                                  &
  \times \delta( \omega - \Omega_{\lambda'} - \Omega_{\lambda''}).
\end{align}
Note that the weighted JDOS is independent from the band index. In
Eq.~(\ref{eq:wjdos-sum}), $N_2^{(1)}(\mathbf{q}, \omega)$ and
$N_2^{(2)}(\mathbf{q}, \omega)$ mean the contributions from two different
scattering classes as written formally,\cite{Physics-of-phonons}
\begin{equation*}
  \begin{cases}
    \text{class 1:} & (\mathbf{q}, \omega)
    + (\mathbf{q}', \Omega_{\lambda'}) \longrightarrow (\mathbf{q}''
    \Omega_{\lambda''})                                    \\
    \text{class 2:} & (\mathbf{q}, \omega) \longrightarrow
    (\mathbf{q}', \Omega_{\lambda'}) + (\mathbf{q}'', \Omega_{\lambda''}),
  \end{cases}
\end{equation*}
respectively. $N_2^{(1)}(\mathbf{q}, \omega)$ and $N_2^{(2)}(\mathbf{q},
  \omega)$ of KCl and NaCl at 300 K are shown in
Figs.~\ref{fig:KCl-jdos-heatmap} and \ref{fig:NaCl-jdos-heatmap} in a similar
manner to the right panels of Figs.~\ref{fig:KCl-self-energy-heatmap} and
\ref{fig:NaCl-self-energy-heatmap}, respectively. $N_2^{(1)}(\mathbf{q},
  \omega)$ (left panels) show weaker intensities than $N_2^{(2)}(\mathbf{q},
  \omega)$ (right panels) due to the term $(n_{\lambda'} - n_{\lambda''})$ in
Eq.~(\ref{eq:wjdos-n1}). $N_2(\mathbf{q}, \omega)$ exhibit high intensities at
higher frequency domains similar to $\Gamma_{\mathbf{q},\text{LO}}(\omega)$ by
the energy conservation $\delta( \omega - \Omega_{\lambda'} -
  \Omega_{\lambda''})$ in Eq.~(\ref{eq:wjdos-n2}). This is considered as the
main reason why the LO modes show strong anharmonicity. To discuss more
details of the LO-mode spectral shapes,
$\Phi_{\lambda\lambda'\lambda''}^{\tilde{\rho}_{\Phi}}$ is necessary,
since $\Gamma_{\mathbf{q},\text{LO}}(\omega)$ show more q-point dependence than
$N_2(\mathbf{q}, \omega)$. This is attributed to wave-like property of the
three phonon interactions.

\subsection{Disconnections of LO-mode branch on $\Gamma$--L path}

In Figs.~\ref{fig:KCl-sf} and \ref{fig:NaCl-sf}, the LO-mode branches of
$A_{\mathbf{q}\nu}(\omega)$ look disconnected on the $\Gamma$--L paths, though
the disconnection is less clear in NaCl. {In each LO-mode branch,
a pair of peaks of
$A_{\mathbf{q},\text{LO}}(\omega)$ gradually change their intensity ratio as
increasing $u$ of $\mathbf{q}=(u, u, u)$ on the $\Gamma$--L path. The pair of
the peaks of KCl are pointed by small arrows in the panel of $u$=0.123 in
Fig.~\ref{fig:KCl-G-L-self-energy}. The peak at lower frequency side (p1) is a
typical main-peak appearing at $\omega_\text{p1} \sim
  \Omega_{\mathbf{q},\text{LO}} +
  \Delta_{\mathbf{q},\text{LO}}(\omega_\text{p1})$. The other peak at higher
frequency side (p2) reflects spike-like change of
$\Delta_{\mathbf{q},\text{LO}}(\omega)$ near $\omega_\text{p2}$. The frequency
of the peak p2 is higher than $\Omega_{\mathbf{q},\text{LO}}+
  \Delta_{\mathbf{q},\text{LO}}(\omega_\text{p2})$, and the intensity of
$A_{\mathbf{q},\text{LO}}(\omega_\text{p2})$ comes from the high-frequency-side
tail of the bell-shaped function centered at $\Omega_{\mathbf{q},\text{LO}}+
  \Delta_{\mathbf{q},\text{LO}}(\omega_\text{p2})$.
$A_{\mathbf{q},\text{LO}}(\omega)$ sharply decays moving away from
$\omega_\text{p2}$ since the spike-like $\Delta_{\mathbf{q},\text{LO}}(\omega)$
increases $\omega - [\Omega_{\mathbf{q},\text{LO}}+
  \Delta_{\mathbf{q},\text{LO}}(\omega)]$ on both sides of the frequency
$\omega_\text{p2}$.}

{In Fig.~\ref{fig:KCl-G-L-self-energy}, we can see that the
position of the spike-like $\Delta_{\mathbf{q},\text{LO}}(\omega)$ shifts to
the lower frequency side as increasing $u$ of $\mathbf{q}=(u, u, u)$. This is
seen clearly in the left panel of Fig.~\ref{fig:KCl-self-energy-heatmap} as a
line of the spike-like $\Delta_{\mathbf{q},\text{LO}}(\omega)$ extending from
just above the top of the LO-mode branch ($\sim$25 THz) at the $\Gamma$-point.
The curve $\Omega_{\mathbf{q},\text{LO}}$ crosses this line around $u$=0.1. At
$u>0.1$, the peak p2 grows to become another main-peak by increasing $u$. In
this way, the disconnection of the LO-mode branch appears. The line of the
spike-like $\Delta_{\mathbf{q},\text{LO}}(\omega)$ is found on the $\Gamma$--X
path side. However, the disconnection does not appear, since
$\Omega_{\mathbf{q},\text{LO}}$ does not cross the line. Compared with KCl,
the line of the spike-like $\Delta_{\mathbf{q},\text{LO}}(\omega)$ is less
clear for NaCl as shown in Fig.~\ref{fig:NaCl-self-energy-heatmap}.}


\section{Summary}
 {
  In this study, we measured the LO-mode phonon branches of KCl and NaCl using IXS
  at 300 K and calculated their spectral functions using the phonon calculation
  under the SSCHA method to investigate their strong anharmonicity. The spectral
  shapes of the IXS measurements and calculations showed good agreements. From the
  calculated spectral functions, we found the multiple peaks of the LO-mode
  spectra and the disconnections of the LO-mode branches on the $\Gamma$--L paths
  in KCl and NaCl, which were unclear in the IXS measurements due to the limited
  energy resolution. From the calculations, we analyzed the spectral shapes of the
  strongly anharmonic LO-modes and how the disconnections of the LO-mode branches
  appear using the phonon self-energies and renormalized phonon
  frequencies.}

\section*{ACKNOWLEDGMENTS}
This work was supported by MEXT Japan through ESISM (Elements Strategy
Initiative for Structural Materials) of Kyoto University and JSPS KAKENHI Grant
Numbers JP21K04632 and JP21K03424. The synchrotron experiments were performed at
the BL35XU of SPring-8 with the approval of the Japan Synchrotron Radiation
Research Institute (JASRI) (Proposal No. 2017A1042, 2017B1297 and 2018A1298).

\appendix
\section{Phonon self energy}
In Appendices, the phonon
modes $(\mathbf{q},\nu)$ and $(-\mathbf{q},\nu)$ are abbreviated by $\lambda$
and $-\lambda$, respectively. In this Appendix, formulae of the phonon
self-energy used in this study is presented.
Three phonon interaction strength
$\Phi_{\lambda\lambda'\lambda''}^{\tilde{\rho}_{\Phi}}$ is given as
\label{sec:self-energy}
\begin{widetext}
  \begin{align}
    \label{eq:ph-ph-strength}
    \Phi_{\lambda\lambda'\lambda''}^{\tilde{\rho}_{\Phi}} =
    \frac{1}{ \sqrt{N}} \frac{1}{3!}
     &
    \sum_{\kappa\kappa'\kappa''}\sum_{j j' j''}
    W_{\lambda,\kappa j}
    W_{\lambda',\kappa' j'}
    W_{\lambda'',\kappa'' j''}
    \sqrt{\frac{\hbar}{2m_{\kappa}  \Omega_{\lambda}  }}
    \sqrt{\frac{\hbar}{2m_{\kappa'}  \Omega_{\lambda'} }}
    \sqrt{\frac{\hbar}{2m_{\kappa''}  \Omega_{\lambda''}  }}\nonumber \\
     &
    \times \sum_{l'l''}
    \Phi_{0\kappa j,l'\kappa'j',l''\kappa''j''}^{\tilde{\rho}_{\Phi}}
    e^{i\mathbf{q}' \cdot
        (\mathbf{r}_{l'\kappa'}-\mathbf{r}_{0\kappa})} e^{i\mathbf{q}''
        \cdot (\mathbf{r}_{l''\kappa''} - \mathbf{r}_{0\kappa})}
    \times e^{i(\mathbf{q}+\mathbf{q}'+\mathbf{q}'')\cdot
        \mathbf{r}_{0\kappa}} \Delta(\mathbf{q}+\mathbf{q}'+\mathbf{q}''),
  \end{align}
  where $N$ is the number of primitive cells, and $m_\kappa$ and
  $\mathbf{r}_{l\kappa}$ are the atomic mass and position in the primitive cell,
  respectively. The symbol $\Delta(\mathbf{q}+\mathbf{q}'+\mathbf{q}'')$ means 1
  if $\mathbf{q}+\mathbf{q}'+\mathbf{q}''$ is a reciprocal lattice vector,
  otherwise 0. $W_{\lambda,\kappa j}$ is the phonon eigenvector obtained as the
  solution of the dynamical matrix of $\Phi$. The phase factor convention of the
  the phonon eigenvectors is based on the same definition of the dynamical
  matrix as written in Ref.~\onlinecite{phono3py}.

  We write real and imaginary parts of the self-energy as
  \begin{align}
    \label{eq:self-energy}
    \Sigma_\lambda(\omega) = \Delta_\lambda(\omega) - i \Gamma_\lambda(\omega),
  \end{align}
  where
  \begin{align}
    \label{eq:real-self-energy}
    \Delta_\lambda(\omega) = \frac{18\pi}{\hbar^2}
    \sum_{\lambda' \lambda''}
    \bigl|\Phi_{-\lambda\lambda'\lambda''}^{\tilde{\rho}_{\Phi}}\bigl|^2
     &
    \left\{
    \left[ \frac{(n_{\lambda'}+ n_{\lambda''}+1)}{
        (\omega-\Omega_{\lambda'}-\Omega_{\lambda''})_\mathrm{p}}
      - \frac{(n_{\lambda'}+ n_{\lambda''}+1)}{
        (\omega+\Omega_{\lambda'}+\Omega_{\lambda''})_\mathrm{p}}
      \right]
    \right.
    \nonumber                                                 \\
     & + \left[ \frac{(n_{\lambda'}-n_{\lambda''})}{(\omega +
    \Omega_{\lambda'} - \Omega_{\lambda''})_\mathrm{p}}
    - \left. \frac{(n_{\lambda'}-n_{\lambda''})}{(\omega -
    \Omega_{\lambda'} + \Omega_{\lambda''})_\mathrm{p}}
    \right]\right\},
  \end{align}
  and
  \begin{align}
    \label{eq:imag-self-energy}
    \Gamma_\lambda(\omega) = \frac{18\pi}{\hbar^2}
    \sum_{\lambda' \lambda''}
    \bigl|\Phi_{-\lambda\lambda'\lambda''}^{\tilde{\rho}_{\Phi}}\bigl|^2
     &
    \left\{(n_{\lambda'}+ n_{\lambda''}+1)
    \left[ \delta(\omega-\Omega_{\lambda'}-\Omega_{\lambda''})
      - \delta(\omega + \Omega_{\lambda'} + \Omega_{\lambda''}) \right] \right.
    \nonumber                         \\
     & + (n_{\lambda'}-n_{\lambda''})
    \left[\delta(\omega+\Omega_{\lambda'}-\Omega_{\lambda''})
      - \left. \delta(\omega-\Omega_{\lambda'}+\Omega_{\lambda''})
      \right]\right\},
  \end{align}
  respectively.
\end{widetext}

\section{Convergence of SSCHA Helmholtz free energy}
\label{sec:convergence-sscha}
SSCHA Helmholtz free energy is considered as a good measure of convergence of
the force constants. The SSCHA Hermholtz free energy $\mathcal{F}_{\Phi}$ is
written as\cite{Errea-SSCHA-2013,Errea-SSCHA-2014,
  Bianco-SSCHA-2017,Monacelli-SSCHA-2021},
\begin{align}
  \label{eq:sscha-fe}
  \mathcal{F}_{\Phi} = \tilde{F}_{\Phi} -
  \langle{\tilde{V}}_{\Phi}\rangle_{\tilde{\rho}_\Phi} +
  \expval{V}_{\tilde{\rho}_\Phi},
\end{align}
where $\tilde{F}_{\Phi}$ and $\langle{\tilde{V}}_{\Phi}\rangle_{\tilde{\rho}_\Phi}$ are the
harmonic Helmholtz free energy and potential energy, and $\expval{V}_{\tilde{\rho}_\Phi}$ is
the potential energy under $\tilde{\rho}_\Phi$. $\tilde{F}_{\Phi}$ is given as
\begin{equation}
  \label{eq:harm-fe}
  \tilde{F}_{\Phi} = \frac{1}{2} \sum_{\lambda}
  \hbar\Omega_\lambda + k_\mathrm{B} T \sum_{\lambda} \ln
  \bigl[1 -\exp(-\hbar\Omega_\lambda/k_\mathrm{B} T) \bigr].
\end{equation}
$\langle{\tilde{V}}_{\Phi}\rangle_{\tilde{\rho}_\Phi}$ is given by Eq.~(A16) of
Ref.~\onlinecite{Bianco-SSCHA-2017} as
\begin{align}
  \label{eq:harm-pot-num}
  \langle{\tilde{V}}_{\Phi}\rangle_{\tilde{\rho}_\Phi} =
   & \frac{1}{2}
  \sum_{l\kappa j,l'\kappa'j'} \Phi_{l\kappa j,l'\kappa'j'}
  \langle u_{l\kappa j} u_{l'\kappa'j'} \rangle_{\tilde{\rho}_\Phi} \\
  =
   & \frac{1}{2} \sum_{l\kappa j,l'\kappa'j'}
  \frac{\Phi_{l\kappa j,l'\kappa'j'}}{N \sqrt{m_\kappa m_{\kappa'}}}
  \sum_{\lambda}
  \sigma^2_{\lambda}(T)
  \nonumber                                                         \\
  \label{eq:harm-pot-anal}
   & \times
  W_{\lambda,\kappa j}^* W_{\lambda,\kappa' j'}
  e^{i\mathbf{q} \cdot (\mathbf{r}_{l'\kappa'} - \mathbf{r}_{l\kappa})},
\end{align}
where the second equation is derived from the displacement
operator:
\begin{align*}
  u_{l\kappa j} & =
  \left( \frac{1}{Nm_\kappa} \right)^{\frac{1}{2}}
  \sum_{\lambda} Q_{\lambda}
  W_{{\lambda},\kappa j}  e^{i\mathbf{q} \cdot \mathbf{r}_{l\kappa}}
  \\
                & =
  \left( \frac{\hbar}{2Nm_\kappa} \right)^{\frac{1}{2}}
  \sum_{\lambda} \Omega_{\lambda}^{-\frac{1}{2}}
  (\hat{a}_{\lambda} + \hat{a}_{-\lambda}^\dagger)
  W_{{\lambda},\kappa j}  e^{i\mathbf{q} \cdot \mathbf{r}_{l\kappa}}.
\end{align*}

\begin{figure}[ht]
  \begin{center}
    \includegraphics[width=1.0\linewidth]{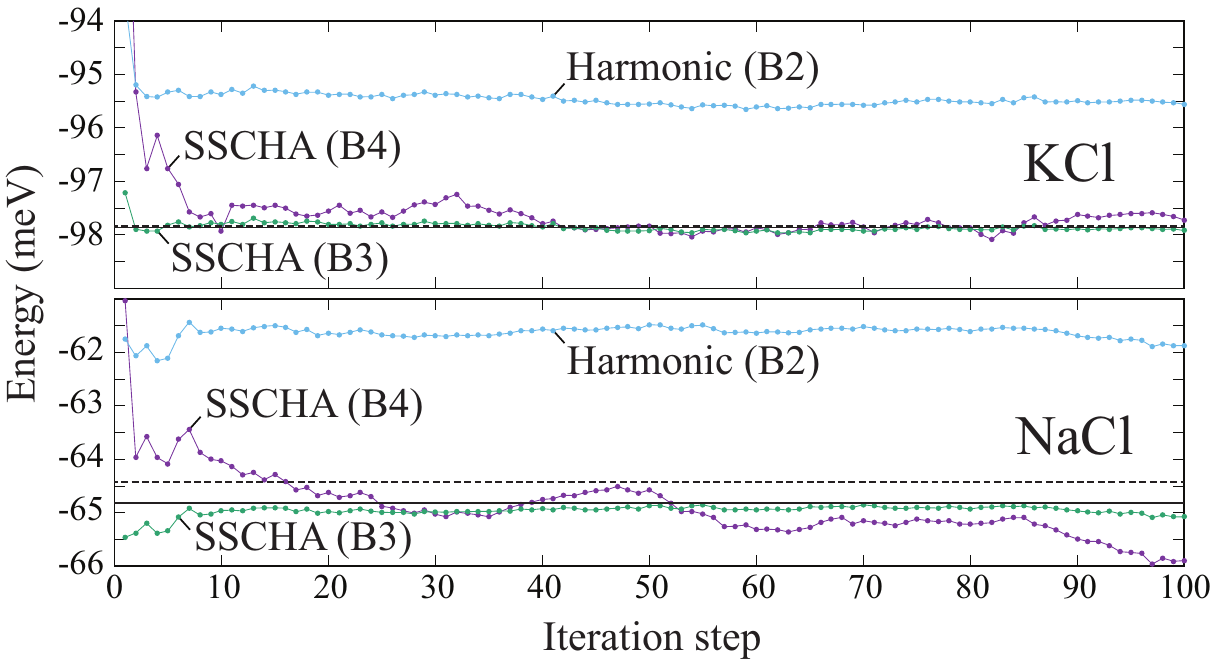} \caption{Energy
      convergences of SSCHA Helmholtz free energies $\mathcal{F}_{\Phi}$
      (Eq.~(\ref{eq:sscha-fe})) of KCl (top) and NaCl (bottom) per primitive
      cells with respect to iteration step. In each figure, the dots labeled by
      SSCHA (B3) and SSCHA (B4) show the $\mathcal{F}_{\Phi}$ in which
      $\langle{\tilde{V}}_{\Phi}\rangle_{\tilde{\rho}_\Phi}$ are computed following
      Eqs.~(\ref{eq:harm-pot-num}) and (\ref{eq:harm-pot-anal}), respectively,
      and those labeled by Harmonic (B2) depict the harmonic Helmholtz free
      energies $\tilde{F}_{\Phi}$ (Eq.~(\ref{eq:harm-fe})). The lines connecting
      the dots are guides to the eye. The solid and dashed horizontal lines in
      each figure are the $\mathcal{F}_{\Phi}$ computed from 4000 supercells
      generated using $\Phi$ obtained at the last iteration step with
      $\langle{\tilde{V}}_{\Phi}\rangle_{\tilde{\rho}_\Phi}$ by Eqs.~(\ref{eq:harm-pot-num})
      and (\ref{eq:harm-pot-anal}), respectively. \label{fig:sscha-convergence}
    }
  \end{center}
\end{figure}

\begin{figure}[ht]
  \begin{center}
    \includegraphics[width=1.0\linewidth]{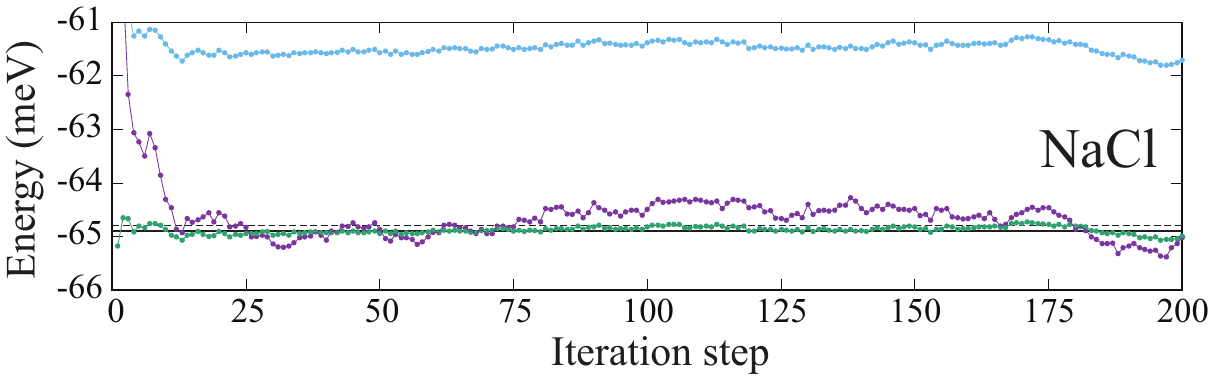} \caption{Energy
      convergences of SSCHA Helmholtz free energies of NaCl. The SSCHA iteration
      was performed twice longer than that in
      Fig.~\ref{fig:sscha-convergence-long}. The points and lines show
      values as explained in Fig.~\ref{fig:sscha-convergence}.
      \label{fig:sscha-convergence-long}
    }
  \end{center}
\end{figure}

Figure ~\ref{fig:sscha-convergence} shows the calculated values of
$\mathcal{F}_{\Phi}$ of KCl and NaCl at iteration steps using the $2\times
  2\times 2$ supercells. In Eq.~(\ref{eq:sscha-fe}), $\expval{V}_{\tilde{\rho}_\Phi}$ were
obtained from electronic total energies of the supercells with generated finite
displacements actually used to compute $\Phi$ relative to those energies without
displacements. $\langle{\tilde{V}}_{\Phi}\rangle_{\tilde{\rho}_\Phi}$ were calculated in two
ways as given in Eqs.~(\ref{eq:harm-pot-num}) and (\ref{eq:harm-pot-anal}). The
values of the former were computed from the displacements same as those used for
the computation of $\expval{V}_{\tilde{\rho}_\Phi}$, and those of the latter were calculated
from phonon frequencies and eigenvectors of $\Phi$. We can see
$\mathcal{F}_{\Phi}$ with $\langle{\tilde{V}}_{\Phi}\rangle_{\tilde{\rho}_\Phi}$ of
Eq.~(\ref{eq:harm-pot-num}) is a more stable measure than $\mathcal{F}_{\Phi}$
with $\langle{\tilde{V}}_{\Phi}\rangle_{\tilde{\rho}_\Phi}$ of Eq.~(\ref{eq:harm-pot-anal}).
The difference between them was found to be smaller in KCl than in NaCl. As the
references, $\mathcal{F}_{\Phi}$ were also computed using 4000 supercells
generated from $\Phi$ obtained at the last iteration step for each of KCl and
NaCl. In KCl, those with $\langle{\tilde{V}}_{\Phi}\rangle_{\tilde{\rho}_\Phi}$ of
Eqs.~(\ref{eq:harm-pot-num}) and (\ref{eq:harm-pot-anal}) are equivalent within
the energy range of the figure, and in NaCl, they are distinguishable. This may
indicate more iteration steps are needed for NaCl than KCl. Another attempt of
NaCl calculation with 200 iteration steps was performed and the result is
presented in Fig.~\ref{fig:sscha-convergence-long}. Difference between
$\mathcal{F}_{\Phi}$ computed using 4000 supercells from
Eqs.~(\ref{eq:harm-pot-num}) and (\ref{eq:harm-pot-anal}) is smaller than that
in Fig.~\ref{fig:sscha-convergence}. This probably indicates the better
convergence of the force constants. However this minor difference impacted
little on the calculation result of the spectral function shapes.

\bibliography{KCl}
\end{document}